\documentclass[12pt]{iopart}

\usepackage{amssymb}
\usepackage{graphicx}
\usepackage{textcomp}
   \graphicspath{{figures/}}

\begin{document}

\title{Controlled insertion and retrieval of atoms coupled to a high-finesse optical resonator}

\author{M.~Khudaverdyan, W.~Alt, I.~Dotsenko\footnote[1]{Present address: Laboratoire Kastler Brossel,
Ecole Normale Sup\'{e}rieure et Coll\'{e}ge de France, 24 rue
Lhomond, 75231 Paris Cedex 05, France.}, T.~Kampschulte,
K.~Lenhard, A.~Rauschenbeutel\footnote[2]{Present address:
Johannes Gutenberg-Universit\"{a}t, Institut f\"{u}r Physik,
Staudingerweg 7, 55099 Mainz, Germany.}, S.~Reick,
K.~Sch\"{o}rner, A.~Widera and D.~Meschede}

\address{Institut f\"{u}r Angewandte Physik, Universit\"{a}t Bonn, Wegelerstr. 8, 53115 Bonn, Germany}

\ead{mika@iap.uni-bonn.de}
\date{\today}
\begin{abstract}
We experimentally investigate the interaction between one and two
atoms and the field of a high-finesse optical resonator.
Laser-cooled caesium atoms are transported into the cavity using
an optical dipole trap. We monitor the interaction dynamics of a
single atom strongly coupled to the resonator mode for several
hundred milliseconds by observing the cavity transmission.
Moreover, we investigate the position-dependent coupling of one
and two atoms by shuttling them through the cavity mode. We
demonstrate an alternative method, which suppresses heating
effects, to analyze the atom-field interaction by retrieving the
atom from the cavity and by measuring its final state.
\end{abstract}

\maketitle

\tableofcontents

\section{Introduction}

Single neutral atoms have been shown to be excellent carriers of
quantum information \cite{Schrader04}. A great challenge and a key
requirement for the utilization of these systems, e.~g. the
construction of multi-particle entangled states, is the
realization of controlled coherent interactions between two or
more individual atoms. The intriguing properties of such systems
may lead to advances in the fields of quantum information
processing and quantum simulation \cite{Nielsen}.

One possible realization of a well controlled interaction is to
couple trapped atoms to the light field of a high finesse optical
resonator \cite{PhysRevLett.75.3788}, thereby mediating an
interaction between the atoms \cite{Zheng2000, Osnaghi2001}.
Coupled atom-cavity systems in the optical domain have been
investigated in several recent experiments
\cite{Nussmann2005,McKeever03}, including deterministic insertion
of the atoms into the cavity field \cite{Fortier2007}. In all
cases the information about the atom-light interaction has been
inferred from the light field. In contrast, in our work we use
deterministic coupling of a single or few atoms to a high finesse
optical cavity and extract complementary information about the
atom-light interaction from both the light leaking from the cavity
and the internal atomic state.

\begin{figure}[!h]
        \centering
        \includegraphics[width=0.9\textwidth]{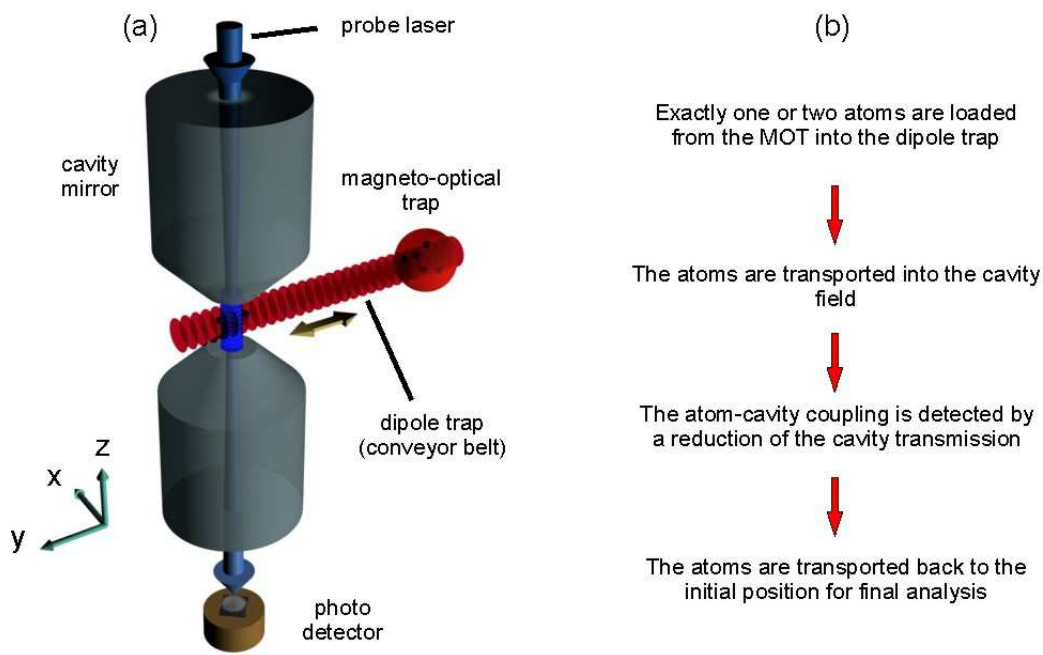}
\caption{Concept of the experiment: (a) schematic drawing of the
experiment and (b) outline of the basic experimental procedure.}
        \label{fig:concept}
    \end{figure}

In our experiment we prepare laser cooled atoms inside a standing
wave dipole trap and deterministically transport them to a
pre-defined position inside the mode of a high-finesse optical
resonator, see figure~\ref{fig:concept}. Due to the strong
atom-cavity interaction, even a single atom causes a well-known
dramatic drop of the observed transmission of a resonant laser
beam (``probe laser") through the cavity \cite{McKeever03,
Puppe2007}. By slowly transporting one or two atoms through the
cavity mode, we map out the spatial variation of the cavity field.
In the experiment with two atoms, we detect an increased effective
coupling compared to the case of a single atom.

Complementing these purely optical measurements, we read out the
final hyperfine state of the atom after its interaction with the
cavity, and thereby we obtain information in a different way. Thus
we pave the way for the detection of entangled or correlated
multi-atom states.

\section{Experimental apparatus}

\subsection{Trapping and transporting single atoms}
\begin{figure}[!t]
        \centering
        \includegraphics[width=0.8\textwidth]{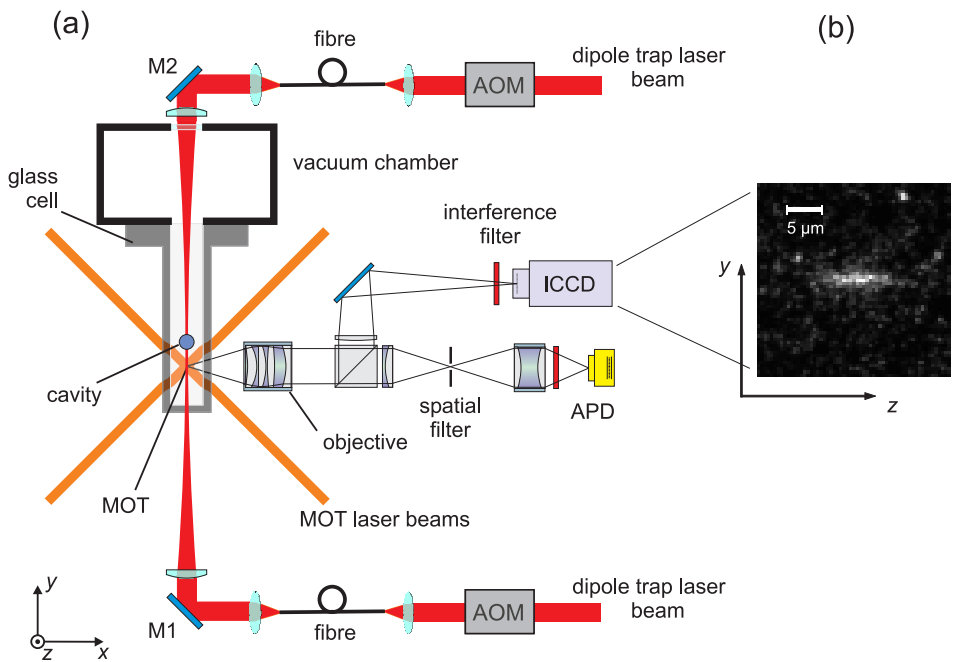}
\caption{(a) Schematic view of the atom trapping setup. Atoms are
loaded from the MOT into the dipole trap and transported along the
dipole trap axis into the cavity mode (oriented along $z$-axis).
The avalanche photodiode (APD) allows us to count trapped atoms,
the ICCD camera provides information on their positions. (b)
Fluorescence image of a single atom, illuminated for 1 second in
the dipole trap.}
        \label{fig:setup}
    \end{figure}

We use a high-gradient magneto-optical trap (MOT) operating inside
an ultra-high vacuum glass cell as a source of cold caesium atoms,
see figure~\ref{fig:setup}. The small capture volume of the MOT
allows us to cool and trap single or few atoms, which are detected
by imaging their fluorescence both onto an intensified CCD camera
and an avalanche photodiode (APD). From the fluorescence signal we
are able to deduce the number of trapped atoms within a few ten
milliseconds.

In order to prepare a predefined number of atoms between 1 and 10,
we use a number-triggered loading technique \cite{L2006}: The MOT
is repeatedly and rapidly loaded until the desired number of atoms
is detected, which are subsequently transferred without loss into
the dipole trap.

The dipole trap is formed by interfering two counter-propagating
laser beams with a wavelength of $1030~$nm. Compared to free space
transmission, a significant improvement  in pointing stability and
beam profile was obtained by using polarization-maintaining
optical fibres with a large mode area (Passive-10/123-PM, LIEKKI)
to guide the laser light from the source to the experimental
setup, see figure~\ref{fig:setup}. These improvements have led to
a reduced scattering from the cavity mirrors, see
section~\ref{stabilization}. The dipole trap laser beams with a
total power of 2.9~W propagate through the MOT and are focused to
a Gaussian beam radius of $w_0=34$~\textmu m, located 1.7~mm away
from the location of the MOT. At the position of the MOT the beam
radius is $w_\mathrm{MOT}=38$~\textmu m and the trap depth equals
$k_\mathrm{B}\times$0.58~mK. Both parameters are inferred from a
measurement of the oscillation frequencies of the atoms in the
trap using the experimental procedure described in
Ref.~\cite{Alt02b}. The measured radial and axial oscillation
frequencies of $\Omega_\mathrm{rad}/2\pi=1.6$~kHz and
$\Omega_\mathrm{ax}/2\pi=262$~kHz, respectively, agree within
10~\% with our calculations based on aberration-free, ideal
Gaussian beams.

In order to image atoms in the dipole trap, we illuminate them
with three-dimensional optical molasses, tuned by several tens of
megahertz to the red of the \mbox{$|F=4\rangle\longrightarrow
|F'=5\rangle$} transition, providing both efficient Doppler
cooling and continuous fluorescence. The position of individual
atoms along the dipole trap axis is determined from the
fluorescence image with an uncertainty of 140~nm
\cite{Dotsenko05}. The storage time of atoms in the dipole trap is
up to several ten seconds.

To transport the atoms along the dipole trap axis into the optical
cavity, acousto-optic modulators (AOMs) are used as frequency
shifters in both beams of the dipole trap. By mutually detuning
their frequencies we can set our standing wave into motion and use
it as an optical conveyor belt. We transport atoms to the position
of the cavity over about 5~mm within 4~ms with sub-micrometer
precision \cite{Dotsenko05}.

\subsection{High-finesse cavity}\label{high-finesse cavity}

\begin{figure}[!t]
        \centering
        \includegraphics[width=0.4\textwidth]{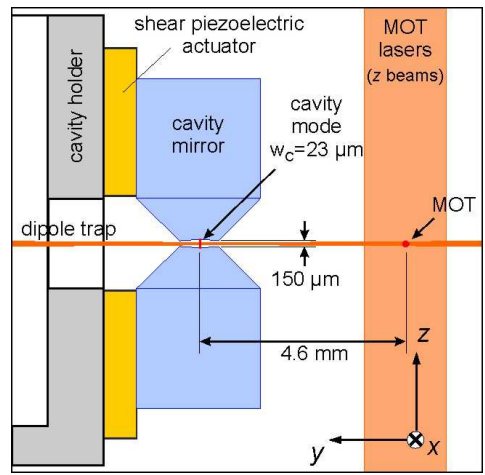}
        \caption{Schematic view of the high-finesse cavity integrated into our atom-trapping
experiment. All elements, laser beams and their spatial
separations are approximately drawn to scale.}
        \label{fig:cavity&traps}
    \end{figure}

Our high-finesse optical cavity is composed of two concave mirrors
consisting of highly reflective, multi-layer dielectric coatings
on super-polished substrates (Research Electro-Optics). The glass
substrates are 3~mm in diameter and are coned to 1~mm at the
mirror surface, see figure~\ref{fig:cavity&traps}. The
reflectivity of each mirror is about 99.9997~\% at the Cs $D_2$
line (852~nm), resulting in a cavity finesse of
$\mathcal{F}\approx1\times10^6$. The transmission of the mirrors
is $T=1.3$~ppm, the losses due to absorption and scattering are
about $A=1.8$~ppm. A large atom-cavity coupling strength is
achieved by realizing a small mode volume with two closely spaced
mirrors. The separation between the mirrors (159~\textmu m) and
their radius of curvature (5~cm) yield a waist radius of the
cavity mode of $w_\mathrm{cav}=23$~\textmu m and result in a mode
volume of about $10^{5}\times\lambda^{3}$. The mirrors are glued
onto shear piezoelectric actuators (PZTs) for tuning the cavity
length and thereby the cavity resonance frequency over 1.5 free
spectral ranges.

\begin{figure}[!t]
        \centering
        \includegraphics[width=0.9\textwidth]{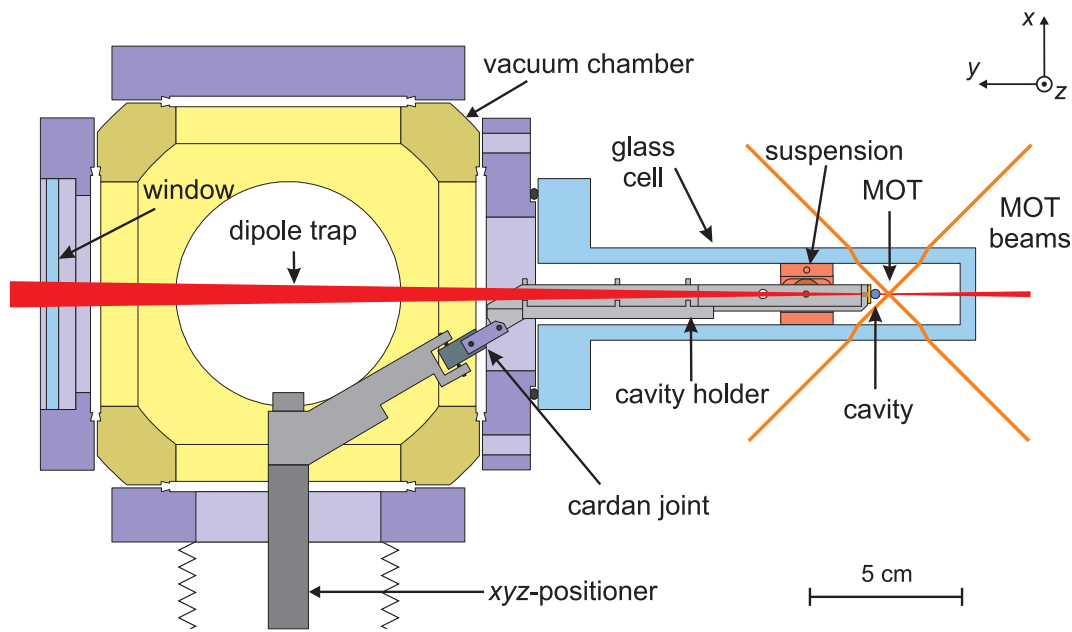}
\caption{Cavity holder in the vacuum setup (top view). The
adjustable holder is used to align the cavity position in the
glass cell relative to the MOT. The resonator is placed about
4.6~mm away from the position of the MOT. }
        \label{fig:cavityHolder}
    \end{figure}

The cavity is assembled on a specially designed cavity holder,
which rests inside the glass cell on a short bellows, see
figure~\ref{fig:cavityHolder}. It is connected via a cardan joint
to a 3D-positioner, consisting of a XY-manipulator (Thermionic
Northwest, XY-B450/T275-1.39) and a Z-feedthrough (Thermionics
Northwest, FLMM133). This combination allows us to adjust the
cavity position with micrometer precision relative to the dipole
trap axis. In order to locate the cavity mode along the dipole
trap axis we transport around 40 atoms having a broad position
distribution along the dipole trap axis towards the cavity mode by
a distance corresponding to an initial guess of the separation
between MOT and cavity mode. In the cavity mode we induce loss of
atoms by heating them with the intra-cavity lock laser. Taking CCD
pictures initially and after transporting back the atoms we
observe the position where atoms have been heated out. By tilting
the dipole trap transverse to its axis, and looking for maximum
atom loss, we can also locate the centre of the cavity mode along
the $x$-axis.

\subsubsection{Atom-light coupling.}
Due to a slight birefringence of our cavity mirrors, the
resonances show a polarization splitting of several linewidths.
The cavity therefore supports linear polarization modes only. For
$\pi$-transitions from $F=4$ to $F'=5$ and for our parameters, the
calculated coupling strength $g/2\pi$ ranges from 8 to 13~MHz for
different $m_\mathrm{F}$ states. These values have to be compared
to the dissipation rates of the coupled atom-cavity system. The
field of the cavity decays at a rate of
$\kappa/2\pi\approx0.4$~MHz, and the atomic dipole decay rate of a
caesium atom is $\gamma/2\pi=2.6$~MHz. Since the condition
$g\gg(\gamma,\kappa)$ is satisfied the system operates in the
\emph{strong coupling regime}. The single atom cooperativity
parameter $C_1=g^2/(2\kappa\gamma)$, quantifying the coherent
energy exchange versus dissipation rates, is expected to be on the
order of 100.

\subsubsection{Dissipation mechanisms.}
The ratio $\kappa/\gamma$ determines the predominant dissipation
mechanism, important for the extraction of information about the
interaction between an atom and the cavity. In the optical domain
often $\kappa \gtrsim  \gamma$ \cite{Nussmann2005, McKeever03,
Fortier2007} and the excitation of the coupled atom-cavity system
predominantly decays via the cavity transmission. However, in our
case $\kappa$ has been made very small in order to achieve a high
$C_1$. Thus, for our value of $\kappa/\gamma=0.15$ the excitation
of the atom-cavity system decays rather by spontaneous emission of
the atom than by the decay of the cavity field. We exploit this
fact by using a spontaneous hyperfine changing Raman transition
and by measuring the final internal atomic state. Such a change in
the internal state can be induced by as few as one to two
spontaneously scattered photons on average. An atom-cavity
interaction can thus be detected even at very short interaction
times, low probe light intensities, and at large laser-atom
detunings by taking ensemble averages. In addition, the detection
efficiency of the atomic state is close to unity in our case, by
far exceeding our photon detection efficiency of only a few
percent.

\subsubsection{Stabilization of the cavity.} \label{stabilization}

Cavity-QED experiments require precise control of the resonance
frequency of the cavity relative to the atomic transition
frequency. In order to keep the cavity resonance frequency stable
within its linewidth, the cavity length must be controlled to
better than $\delta L \leq \lambda/(2\mathcal{F})=0.4$~pm in our
case. During the entire experimental procedure it is therefore
necessary to actively stabilize the cavity length against acoustic
vibrations and thermal drifts. There are two major experimental
issues to be considered: First, there is only marginal vibration
isolation provided by the bellows (see section \ref{high-finesse
cavity}). Second, switching of the laser beams of the dipole trap
causes thermal expansion of the cavity mirror assembly due to
residual absorption. As a consequence, the unstabilized cavity
resonance frequency drifts with an initial rate of about
$2\times10^{5}$~linewidths per second within the first 300~ms
after the dipole trap has been switched on. The corresponding
servo loop therefore uses optimizations such as a double
integrator and a notch filter compensating a mechanical resonance
of the PZTs, and has a bandwidth of about 10~kHz.

We use an auxiliary far blue-detuned laser (``lock laser") at
840~nm for frequency stabilization. We thus suppress excitation of
an atom by the intra-cavity lock laser field. Due to resonant
enhancement the injected lock laser power of 0.4~\textmu W also
creates a blue-detuned standing wave dipole potential with a
height of about 0.3~mK and a maximum scattering rate of
40~$\mathrm{s}^{-1}$.

\begin{figure}[!t]
        \centering
        \includegraphics[width=0.9\textwidth]{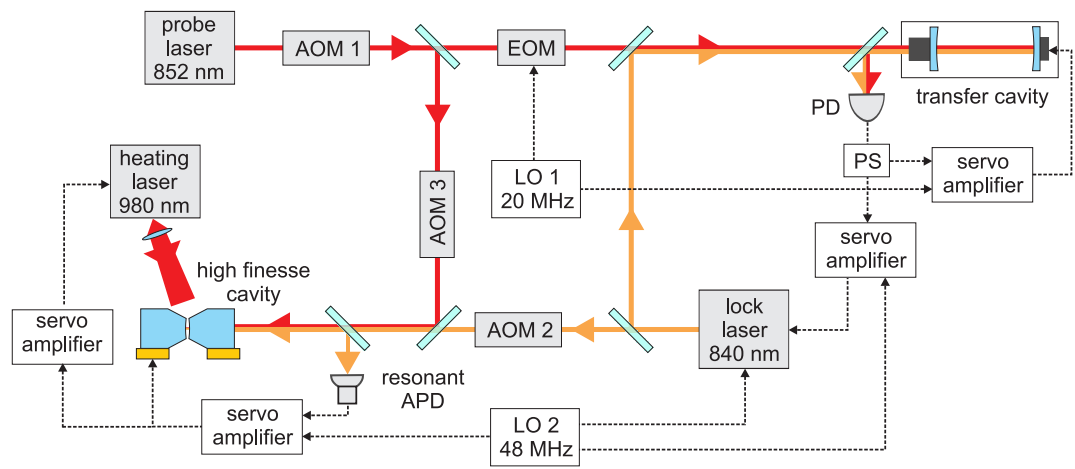}
\caption{Frequency stabilization of the high-finesse cavity. Using
the transfer cavity and the lock laser, the stability of the probe
laser, which is stabilized to a caesium polarization spectroscopy,
is transferred to the high-finesse cavity. AOM: acousto-optic
modulator, EOM: electro-optic modulator, PD: photodiode, APD:
avalanche photodiode, LO: local oscillator, PS: RF power splitter.
The three servo loops are based on the PDH method.}
        \label{fig:stabilization}
    \end{figure}

Our locking scheme for the stabilization of the high-finesse
cavity is similar to the one presented in \cite{Mabuchi99}. Its
main elements are schematically depicted in
figure~\ref{fig:stabilization}. Because of the absence of easily
accessible atomic frequency references at $\lambda=840~$nm, the
lock laser itself is stabilized onto an auxiliary cavity, which
transfers the frequency stability of the probe laser to the lock
laser. The error signals for all servo loops are based on the
Pound-Drever-Hall (PDH) method \cite{Drever83}.
The acousto-optic modulators (AOM 1-3 in
figure~\ref{fig:stabilization}) allow independent control of the
probe-cavity detuning and probe-atom detuning within a range of
$\pm100$~MHz.

Although the cavity resonance at 840~nm is stabilized onto the
lock laser, we still observe drifts of the cavity resonance
frequency at 852~nm with respect to the probe laser frequency on a
timescale of several seconds. This residual frequency deviation is
caused by different temperature dependencies of the effective
penetration depths into the mirror coatings for the two
wavelengths. In order to compensate for this differential drift,
we scan the cavity resonance over the probe laser frequency using
AOM~2 and record a transmission spectrum of the cavity. Then, we
determine the AOM control voltage which corresponds to the maximum
transmission of the probe laser, and use this value in the
following experimental cycle.

In order to compensate for large changes of the cavity length
caused by long term temperature variations exceeding the tuning
range of the piezoelectric actuators, we heat the cavity  using a
multimode ``heating laser" at 980~nm and a power of up to 400~mW.
Since we cannot directly measure the temperature of the cavity
mirrors, we use the voltage applied to the piezoelectric actuators
as an error signal. This signal is fed back onto the power of the
heating laser in a slow servo loop, thus keeping the PZT offset
voltage close to zero.

\subsubsection{Detection of the cavity transmission.}

Both the lock laser and the probe laser beams share the same
transverse mode profile. For the effect of the different
longitudinal modes see section~\ref{ch:trapping dynamics} and
figure~\ref{fig:TransmissionSignalJumps}. Since information about
the atom-light interaction is only contained in the probe beam
transmission, we separate them at the output of the cavity by a
diffraction grating. For the detection of the probe beam
transmission we use a fiber-coupled single photon counting module
(SPCM). It has a total measured quantum efficiency of about $30\%$
and a dark count rate of 500~counts/s. The detection efficiency,
i.~e.~the power detected by the SPCM divided by the power of the
probe beam directly at the output of the cavity, is approximately
$9\%$, including the diffraction efficiency of the grating,
transmission of an interference filter, the losses on the
remaining optics, and the quantum efficiency of the SPCM. In
addition, since the light is partially absorbed and scattered by
the cavity mirrors, the total efficiency for detection of
intra-cavity photons is further reduced by a factor of
$T/[2(T+A)]$ to about 2~\%.

\section{Strong coupling of a single atom to the resonator mode} \label{ch:Strong Coupling}

In order to characterize the interaction between atoms and the
field mode, we have first carried out experiments with single
atoms. We always tune the cavity into resonance with the probe
laser, i.~e. $\Delta_\mathrm{c}=\Delta_\mathrm{p}$, where
$\Delta_\mathrm{c}=\omega_\mathrm{c}-\omega_\mathrm{a}$ and
$\Delta_\mathrm{p}=\omega_\mathrm{p}-\omega_\mathrm{a}$. Here,
$\omega_\mathrm{p}$, $\omega_\mathrm{c}$, and $\omega_\mathrm{a}$
are the angular frequencies of the probe laser, the cavity
resonance, and the AC-Stark shifted atomic $|F=4\rangle
\longrightarrow |F'=5\rangle$ transition, respectively. The
spatial profile of the coupling strength $g$ as well as the
spatial distribution of the probe laser field inside the cavity
are defined by the $\mathrm{TEM}_{00}$ cavity mode. Upon the
insertion of a single atom into the resonator mode the initially
high transmission of a resonant probe laser drops due to the
interaction-induced normal mode splitting.

\subsection{Coupling to the centre of the mode}\label{Coupling of a single atom to the centre of the cavity mode}

  \begin{figure}[!t]
        \centering
        \includegraphics[width=1\textwidth]{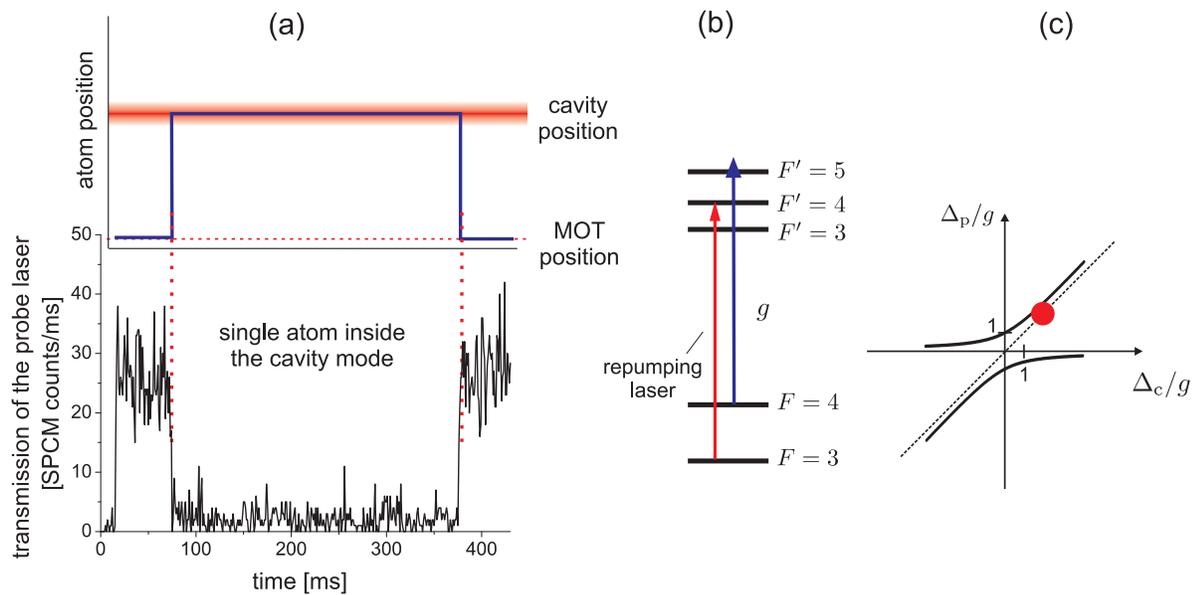}
\caption{(a) Transmission signal versus time obtained in a single
experimental run showing the case of continuous strong coupling.
Upon placement of a single atom into the cavity mode the
transmission drops to 5~$\%$ of its value for the empty cavity.
After continuously observing the atom for 300~ms inside the cavity
it is transported back to the position of the MOT. (b) In this
measurement the cavity resonance is blue detuned to the
$|F=4\rangle\longrightarrow |F'=5\rangle$ transition by
$\Delta_\mathrm{c}/2\pi\approx24$~MHz. (c) Schematic illustration
of the energies of the coupled atom-cavity states (solid lines)
and empty cavity state (dashed line). Our detunings
$\Delta_\mathrm{c}$ and $\Delta_\mathrm{p}$ are shown as a red
circle.}
        \label{fig:noTransmissionSignal}
    \end{figure}

In order to demonstrate the coupling of a single atom to the
cavity centre, we monitor the transmission of the probe laser beam
while inserting the atom. Figure~\ref{fig:noTransmissionSignal}
presents an example of a single experimental cycle. At time $t=0$
the SPCM is switched on. The background signal of about
$10^3$~counts/s corresponds to the dark counts and the stray light
detected by the SPCM. The cavity and the probe laser are blue
detuned by $\Delta_\mathrm{c}/2\pi\approx24$~MHz. There exists
theoretical work which indicates cooling in this regime
\cite{Domokos2003}. However, we found this value empirically by
observing an increased lifetime of the atoms inside the cavity.

The probe laser beam is switched on at $t =14$~ms resulting in a
transmission count rate of $26\times10^3$~counts/s, corresponding
to a mean intra-cavity photon number of  about 0.1, indicating
that we are in the regime of weak atomic excitation\footnote[1]{At
this intra-cavity photon number for an empty cavity the scattering
rate of an atom coupled to the cavity is much smaller than
$\gamma$ for any detuning $\Delta_\mathrm{c}$ and any coupling
strength $g$.}. At $t=70$~ms we place an atom into the centre of
the cavity mode, which causes the transmission of the probe laser
to drop to $2\times10^3$~counts/s. Subtracting the background
count rate, the transmission drops down to approximately $5~\%$ of
its initial level. For our values of $\kappa$ and $\gamma$ and the
chosen value of $\Delta_\mathrm{c}$, the observed drop in the
probe transmission indicates strong coupling of the atom to the
field of the resonator mode. During the entire experimental cycle
a repumping laser resonant with the $|F=3\rangle\longrightarrow
|F'=4\rangle$ transition is applied along the dipole trap axis. It
transfers the atom back into the $F=4$ state if it is
off-resonantly pumped into the $F=3$ state by the probe laser. The
transmission remains at this level of 5~$\%$ while the atom
resides in the cavity until we transport the atom back to the
position of the MOT. There we can take a second picture of the
atom to not only verify its presence after the interaction with
the cavity, but also to check that its position in the standing
wave dipole trap has not changed, i.~e. it has not been
temporarily heated out of its potential well.

\begin{figure}[!t]
        \centering
        \includegraphics[width=0.8\textwidth]{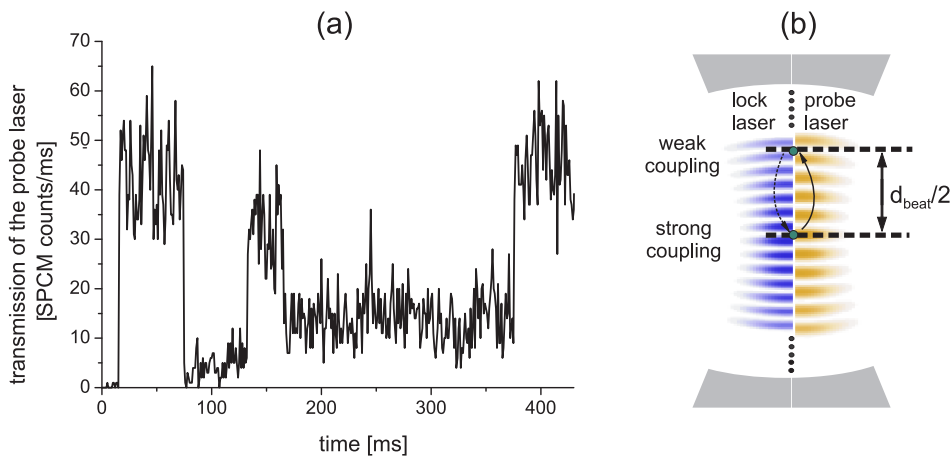}
\caption{(a) Transmission signal of the probe laser beam through
the cavity for the case of time-dependent coupling. (b) Due to
jumps between different nodes of the repulsive lock laser standing
wave, the coupling strength of the atom to the cavity changes.}
        \label{fig:TransmissionSignalJumps}
    \end{figure}

\subsubsection{Model.}
In order to quantitatively analyze our measurements and to
establish a relation between $g$ and the observed average
transmission, we model our physical system including a variation
of $g$ caused by internal and external dynamics of the atom. For
this purpose we numerically solve the master equation of the
coupled atom-cavity system \cite{Carmichael}. All processes
causing a variation of the coupling strength $g$ are slow compared
to the decay time 1/$\kappa$ of the cavity field. Therefore, the
cavity transmission at any moment is determined by the
instantaneous coupling strength. Since $g$ is time-dependent and
the cavity transmission is a non-linear function of $g$ we
simulate the measured average transmission level by first
calculating the transmission levels corresponding to all possible
values of $g$, and then computing the weighted average over these
levels.

The internal dynamics of the atom is caused by its continuous
scattering of photons while coupled to the cavity mode, which
causes changes in the Zeeman sublevel occupation. Since the atom
is not optically pumped into a specific $m_\mathrm{F}$ sublevel of
the ground state $F=4$, we assume a homogeneous distribution over
all these states and average over the transmissions corresponding
to the different $m_\mathrm{F}$ states.

The external dynamics is given by the motion of the atom in the
trapping potential. We assume a thermal Boltzmann energy
distribution neglecting modifications due to cavity QED effects,
the multi-level atomic structure and technical noise which could
lead to a non-thermal energy distribution \cite{Domokos2003}.
Since our integration time is much longer than the heating and
cooling timescales, our model allows to assign an effective
temperature corresponding to the time averaged energy $\langle E
\rangle$ of the atom in the trap via $\langle E
\rangle=3k_\mathrm{B}T$. We take into account the variation of $g$
and the variation of the AC-Stark shift, due to the dipole trap
and the lock laser, caused by the oscillatory motion. The depth of
the dipole potential due to the intra-cavity photon number of the
probe laser corresponding to the empty cavity ($n=0.1$) is only
about 30~\textmu K at the centre of the cavity mode
\cite{Brune1994}. An atom coupled to the mode strongly reduces the
photon number which further decreases this potential to values
much lower than the atomic temperature and all other trapping
potentials. We therefore neglect the effect of this potential.

Along the $x$-axis, transverse to both dipole trap and cavity
axis, the atom is weakly confined by the Gaussian profile of the
dipole trap. Due to oscillations along this direction the atom
experiences a variation of the AC-Stark shift and the coupling
strength $g$. Along the $z$-axis the atom is well localized to
several hundred nanometers by the lock laser standing wave with a
potential height of $U/k_\mathrm{B}\approx0.3~\mathrm{m}$K. Along
the dipole trap axis ($y$-axis) the variation of $g$ is negligible
and we consider the variation of the AC-Stark shift only.

We use the model temperature as a fit parameter in our numerical
simulations and find that a temperature of about 0.17~mK
reproduces our observed drop of the transmission to 5~\%. This
result indicates a somewhat higher temperature than the typical
0.07~mK inside our dipole trap after molasses cooling possibly due
to different cooling and heating mechanisms present inside the
cavity.

\subsubsection{Observation of trapping dynamics.} \label{ch:trapping dynamics}
In experimental records we frequently observe strong variations of
the transmission as shown in
figure~\ref{fig:TransmissionSignalJumps}(a). We attribute these to
hopping of atoms between different trapping sites along the cavity
axis (\mbox{$z$-axis}): Due to the dynamic equilibrium between
cooling and heating processes in the cavity, the atom can be
heated out of one node of the blue-detuned lock laser standing
wave and subsequently be cooled into a different node. Since the
lock and probe laser standing waves have different periodicities
due to their different wavelengths, the coupling strength at
potential minima of the lock laser standing wave changes from
maximum to minimum over a length of $d_\mathrm{beat}/2=15$~\textmu
m, as depicted in figure~\ref{fig:TransmissionSignalJumps}(b).
Therefore, the hopping of atoms along the $z$-axis can result in
sudden strong changes of the atom-cavity coupling strength,
leading to jumps of the transmission level. We observe that the
rate of these transmission variations seems to depend critically
on a complex interplay between lock and probe laser intensities,
detuning, and the initial insertion conditions. This phenomenon,
however, is not yet fully understood and requires further
investigation.

Note that the data presented in
figure~\ref{fig:noTransmissionSignal} have been selected for
strongest coupling and no hopping. In the corresponding
simulations we have assumed the atom to reside at a potential
minimum coinciding with a maximum of $g$.

\subsection{Position-dependent coupling}
  \begin{figure}[!t]
        \centering
        \includegraphics[width=0.95\textwidth]{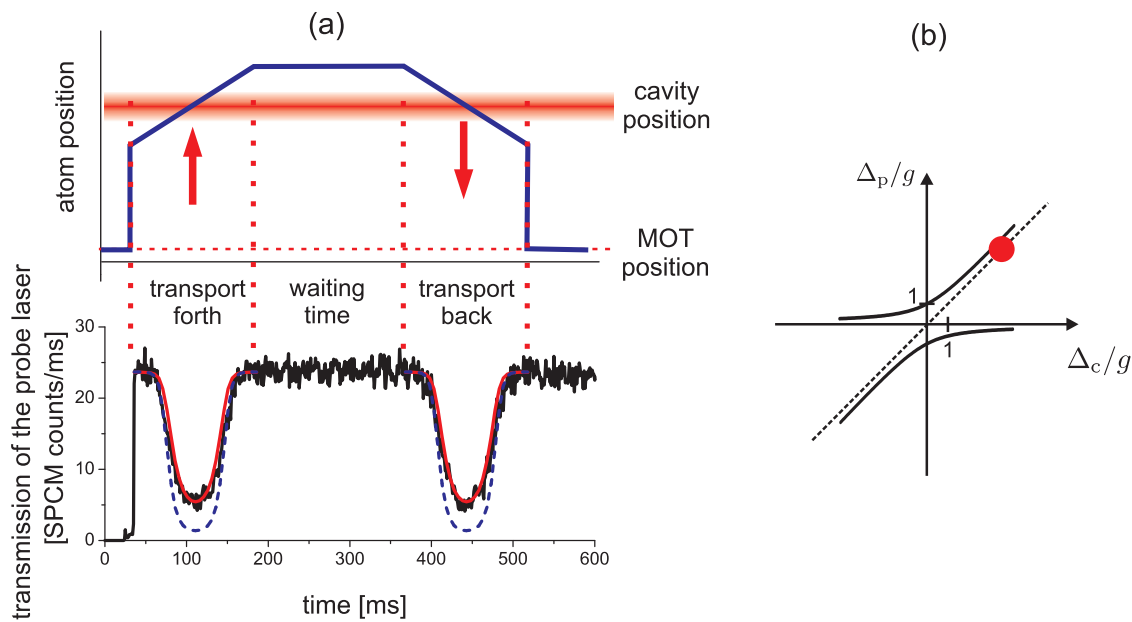}
\caption{(a) Transmission of the probe laser beam as an atom is
slowly swept forth and back across the cavity mode over a distance
of $100$~\textmu m within 150~ms. The trace is an average over 19
experimental runs. The measured data are in a good agreement with
the numerical simulation (solid line in red). The probe laser is
is blue-detuned with respect to the $|F=4\rangle \longrightarrow
|F'=5\rangle$ atomic transition by
$\Delta_\mathrm{c}/2\pi\approx44$~MHz. Dashed blue line indicates
the cavity transmission assuming no thermal motion and strongest
coupling along the cavity axis. (b) Schematic illustration of the
energies of the coupled atom-cavity states, our detunings
$\Delta_\mathrm{c}$ and $\Delta_\mathrm{p}$ are shown as a red
circle.}
        \label{fig:sweepingAtom}
    \end{figure}

Applications of cavity QED in quantum information experiments with
a string of atoms usually require control of the coupling
strengths for different atoms. This can be achieved e.~g. by
suitably positioning them inside the mode profile. In order to map
out the spatial distribution of the coupling strength transverse
to the cavity axis we continuously observe the transmission while
transporting a single atom slowly through the mode, similar to the
experiments reported in \cite{Nussmann2005, Fortier2007}.

We first shuttle the atom to the edge of the mode $50$~\textmu m
from the mode centre, where $g$ is negligible. As it is depicted
in figure~\ref{fig:sweepingAtom}, by slowly transporting the atom
over 100~\textmu m across the cavity mode within 150~ms, we
observe a continuous variation of the transmission caused by a
variation of $g$. After a waiting time the direction of the slow
transport is reversed. For both transportation directions the
average transmission drops to about $20~\%$ of its maximum level
in the centre of the cavity mode, and the shape of the
transmission is almost identical.

The drop to about 20~\% is less than the reduction in the previous
single shot experiment. This is a result of averaging over 19
shots, selected only for the presence of the atom during the
complete transportation in both directions, and of a larger blue
detuning of $\Delta_\mathrm{c}/2\pi\approx44~$MHz. Due to thermal
oscillations along the $z$-axis inside the dipole trap, the atom
enters the cavity field in different nodes of the lock laser
standing wave, thus causing the coupling strength and the
reduction of the transmission to have different values for each
transport through the cavity mode. We perform a numerical
simulation to compare our observations with the prediction of our
simple model. With the temperature as a single fit parameter, our
numerical simulations are in good agreement with the measured data
by assuming a temperature of about 0.13~$\mathrm{m}$K.

\section{Coupling of one and two atoms}  \label{ch:Atom-number dependent coupling stregth }
\begin{figure}[!t]
        \centering
        \includegraphics[width=0.5\textwidth]{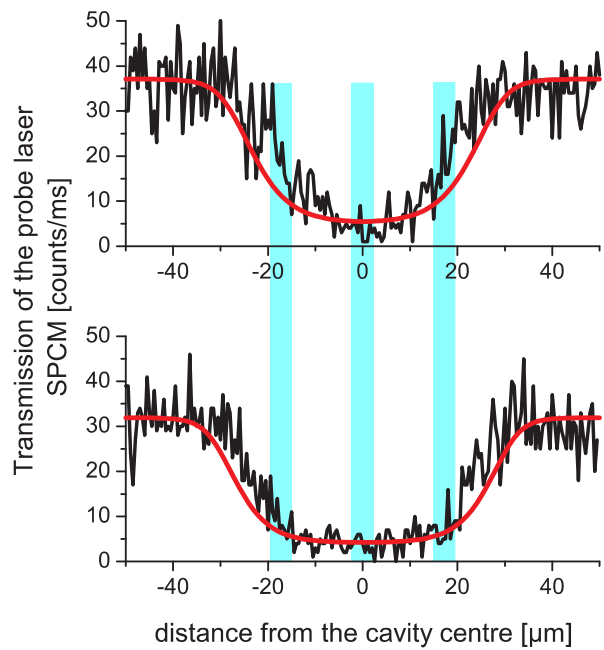}
\caption{Transmission of the probe laser beam as a single atom (on
the top) and two atoms (below) are swept across the cavity mode.
For the latter case the width of the transmission dip is notably
larger, indicating a stronger coupling strength. The probe laser
is is blue-detuned with respect to the $|F=4\rangle
\longrightarrow |F'=5\rangle$ atomic transition by
$\Delta_\mathrm{c}/2\pi\approx24$~MHz. The measured data agrees
reasonably well with the numerical simulation (solid red lines)
for one and two atoms in the upper and lower graph,
respectively.}\label{fig:sweeping1&2Atoms}
    \end{figure}

For \emph{N} atoms simultaneously coupled to the cavity mode the
effective collective coupling strength $g_N$ is expected to scale
as $g\sqrt{N}$ for weak excitation, where $g$ is the coupling
strength of a single atom. In figure~\ref{fig:sweeping1&2Atoms} we
compare the cavity transmission when sweeping either one or two
atoms slowly through the cavity mode. In the case of two atoms
their separation is below 1.5~\textmu m, which is small compared
to the waist of the cavity mode. We therefore assume that the two
atoms are interacting with the cavity mode simultaneously at the
same position. The data in figure~\ref{fig:sweeping1&2Atoms} are
single experimental runs and are selected for strongest coupling
along the $z$-axis, as discussed in the previous section. In the
central region of the cavity already a single atom almost
completely blocks the transmission of the probe laser. However, at
the outer regions of the cavity mode two atoms cause a
significantly stronger drop in transmission than a single atom.

In analogy to the previous section, we performed a numerical
simulation, assuming that the intra-cavity temperature of the
atoms is the same as in the experiment presented in
figure~\ref{fig:noTransmissionSignal}. The case of two atoms is
approximated here by a single atom coupled to the cavity with a
strength $g\times\sqrt{2}$. The numerical simulation agrees well
with the measured data for the central region of the cavity. In
the outer regions, both measured transmission levels are slightly
higher than the calculation. This deviation is to be expected, as
our simple model assumes constant confinement of the atom by the
blue-detuned lock-laser standing wave independent of the position
along the dipole trap axis. However, this confining potential
decreases towards the edge of the cavity mode, and at some point
the atom is no longer confined by the lock laser standing wave. As
a result, the atom oscillates freely along the $z$-axis and
experiences on average a weaker coupling strengths, thus
increasing the transmission level.

\section{Detection of the final atomic state} \label{Detection of atomic
state}

We now present a complementary method to extract information about
atom-cavity coupling which is based on the detection of the
internal atomic state after an atom has interacted with the field
of the resonator. In the microwave domain this method is used in
the cavity QED experiments of S.~Haroche, J.-M.~Raimond and
co-workers \cite{Gleyzes2007}.

For this measurement, we prepare a single atom in the dipole trap
in the hyperfine ground state $F=4$ by illuminating it with the
MOT repumping laser. We then transport the atom to the cavity mode
and subsequently inject the probe laser beam into the cavity. In
this experiment the cavity is blue detuned by about 40~MHz from
the \mbox{$|F=4\rangle\longrightarrow |F'=3\rangle$} transition
and is red detuned by 160~MHz from the \mbox{$|F=4\rangle
\longrightarrow |F'=4\rangle$} transition. During the interaction
time of 1~ms the injected probe laser power corresponds to about
0.04 photons inside the cavity, much less than what was used in
the previous experiments presented above. The atom is pumped into
the non-interacting $F=3$ ground state with a probability
proportional to the intra-cavity probe light intensity, after
scattering a few photons only. We retrieve the atom from the
cavity and subsequently detect its final state with an efficiency
close to unity by applying a ``push-out" laser \cite{Kuhr03}. This
laser is resonant with the $|F=4\rangle \rightarrow |F'=5\rangle$
transition and removes all atoms in $F=4$ from the trap.
Subsequent transfer of the remaining atoms back to the MOT reveals
the number of atoms in $F=3$. In order to ensure the efficient
operation of the ``push-out" laser we reduce the depth of the
dipole trap after transporting atoms back to the MOT position,
which limits the overall survival probability of the atoms in
$F=3$ to $77_{-9}^{+7}~\%$.

 \begin{figure}[!t]
        \centering
        \includegraphics[width=0.8\textwidth]{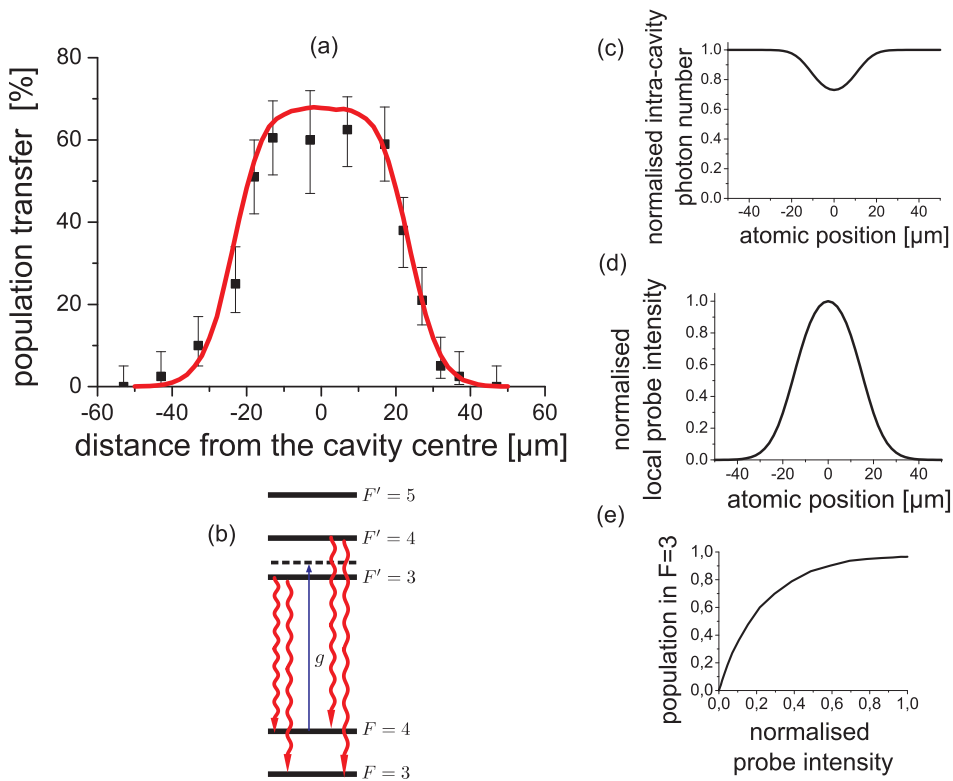}
        \caption{(a) Measurement of the final state of the atom after it has interacted with
        the field of the cavity mode.
        Each point is the result of about
        40 experimental runs with a single
        atom. Model for an atom in the state $|F=4, m_\mathrm{F}=1\rangle$: (c) intra-cavity photon number and (d) local probe laser
        intensity as a function of the atomic position and (e) pumping efficiency as a function of local probe
        laser intensity. \\ The solid red line in (a) is then obtained by taking the composition of
        (d) and (e), performing these calculations for each state $|F=4, m_\mathrm{F}\rangle$, and averaging subsequently. (b) The probe laser is blue detuned from the $|F=4\rangle \longrightarrow |F'=3\rangle$
        transition by 40~MHz and is red detuned from the $|F=4\rangle \longrightarrow |F'=4\rangle$ transition by 160~MHz.}
        \label{fig:modePosition}
    \end{figure}

For each position of the atom inside the cavity mode we take an
ensemble average of about 40 repetitions. By varying the atomic
position across the cavity mode, we map out the optical pumping
rate due to the probe laser beam inside the cavity, see
figure~\ref{fig:modePosition}(a). We observe a peak in population
transfer corresponding to the transverse cavity mode profile with
a maximum transfer efficiency of about $60~\%$. In this experiment
an atom interacts dispersively with the cavity field and has only
a small effect on the intra-cavity photon number. Therefore this
photon number still weakly depends on the distance of the atom
from the cavity mode centre and on the initial $m_\mathrm{F}$
state of the atom. As an example, figure~\ref{fig:modePosition}(c)
shows the normalized photon number as a function of atomic
distance from the cavity centre in the case of an atom occupying
the state $|F=4, m_\mathrm{F}=1\rangle$. The resulting probe laser
intensity, shown in figure~\ref{fig:modePosition}(d), is slightly
broader than the Gaussian intensity profile in the absence of an
atom. Compared to figure~\ref{fig:modePosition}(d), the transfer
efficiency curve is further broadened due to the nonlinear
dependence of the transfer efficiency on the probe light
intensity. This dependence, shown in
figure~\ref{fig:modePosition}(e), is modelled for each initial
$m_\mathrm{F}$ state by performing Monte-Carlo simulations and
taking thermal motion into account, as was described in
section~\ref{Coupling of a single atom to the centre of the cavity
mode}. We divide the holding time of an atom inside the cavity
into 10~\textmu s-intervals, short compared to the average
scattering rate. In each of these intervals the atom is assumed to
be either excited with a probability determined by the local probe
laser intensity and the corresponding Clebsch-Gordan coefficient
and then to decay to the hyperfine state $F=4$ or $F=3$ with the
corresponding branching ratios; or it remains in the state $F=4$.
The result of this simulation for the initial state $|F=4,
m_\mathrm{F}=1\rangle$ is shown in
figure~\ref{fig:modePosition}(e). It is used to compute the
spatial variation of the population transfer from the probe
intensity distribution of figure~\ref{fig:modePosition}(d). We
again assume a homogeneous distribution over different
$m_\mathrm{F}$ states and average the population transfer over
these states. Since the simulation shows only a very slight
dependence on the temperature, we use the intra-cavity photon
number as a single fit parameter. The resulting photon number of
0.02 equals 60~\% of the expected photon number estimated from the
transmission. We attribute this deviation to systematic
uncertainties of our model and inaccuracies of our measurement.
The theoretical curve in figure~\ref{fig:modePosition}(a) is
obtained by normalizing the simulated averaged population transfer
to the survival probability of 77~\% and fits well to the observed
data.

\section{Summary and Outlook}

We have presented an apparatus to strongly couple a
well-controlled number of atoms to a high-finesse optical
resonator. Observation of the cavity transmission could allow us
to detect and examine the motional dynamics of one and two atoms
in the cavity \cite{Domokos2003,Asboth05}. The observed
position-dependent variation of the interaction strength, together
with an improved localization of atoms inside the cavity offers a
tool to fully control the atom-cavity coupling strength. Measuring
the final hyperfine state of the atom allows us to detect the
interaction with the cavity while heating effects are strongly
suppressed due to the short interaction time and small probe laser
power. Using this method, the next step is to reveal the coherent
nature of the atom-cavity interaction by observing the vacuum Rabi
splitting \cite{Sanchez-Mondragon1983, Boca2004,Maunz2005} once we
have established a better control over the coupling strength. We
finally aim at the generation of entanglement between two atoms
and at the creation of multi-atom correlated quantum states
utilizing our ability to selectively initialize and manipulate the
internal state of individual atoms \cite{Schrader04}.

\section*{Acknowledgment}
We thank A~Stiebeiner for her help with the cavity stabilization.
This work was in early stages supported by the Deutsche
Forschungsgemeinschaft within the Schwerpunkt
``Quanten-Informationsverarbeitung" as well as by the European
Commission within the project ``QGATES". SR acknowledges support
from the ``Deutsche Telekom Stiftung" and TK acknowledges support
from the ``Studienstiftung des deutschen Volkes" and from the
``Bonn-Cologne Graduate School of Physics and Astronomy".

\end{document}